\documentclass[final,3p,times]{elsarticle}

\usepackage{amssymb}

\makeatletter
\def\Hline{%
\noalign{\ifnum0=`}\fi\hrule \@height 1.5pt \futurelet
\reserved@a\@xhline}
\makeatother

\journal{Journal of Computational Physics}

\begin{document}

\begin{frontmatter}



\title{A hyperbolic-equation system approach for magnetized electron fluids in quasi-neutral plasmas}


\author[label1]{Rei Kawashima\corref{cor1}}

\cortext[cor1]{Corresponding author. Tel.: +81 3 5841 6559; fax: +81 3 5841 6559.}

\address[label1]{Department of Aeronautics and Astronautics, The University of Tokyo, 7-3-1 Hongo, Bunkyo-ku, Tokyo 113-8656, Japan}

\author[label2]{Kimiya Komurasaki}
\address[label2]{Department of Advanced Energy, The University of Tokyo, 5-1-5 Kashiwanoha, Kashiwa, Chiba 277-8561, Japan}

\author[label1]{Tony Sch\"{o}nherr}

\begin{abstract}
A new approach using a hyperbolic-equation system (HES) is proposed to solve for the electron fluids in quasi-neutral plasmas. The HES approach avoids treatments of cross-diffusion terms which cause numerical instabilities in conventional approaches using an elliptic equation (EE).
A test calculation reveals that the HES approach can robustly solve problems of strong magnetic confinement by using an upwind method. The computation time of the HES approach is compared with that of the EE approach in terms of the size of the problem and the strength of magnetic confinement. The results indicate that the HES approach can be used to solve problems in a simple structured mesh without increasing computational time compared to the EE approach and that it features fast convergence in conditions of strong magnetic confinement. 
\end{abstract}

\end{frontmatter}



\section{Introduction}
Electron fluids in quasi-neutral plasmas appear in many practical plasma simulations such as space propulsion, astrophysics, and plasma processing \cite{Parra,kajimura2010hybrid,GRL:GRL8860,Wang}. In simulations of quasi-neutral plasmas, one commonly uses the plasma approximation, in which the space potential is solved using the electron conservation equations, rather than by using Gauss's law \cite{chen1984introduction}. In conventional methods, the equations of conservation of mass and conservation of momentum are integrated into one elliptic equation (EE), and this equation is solved for the space potential as a boundary value problem. However, this EE is an ``anisotropic diffusion equation" and, owing to magnetic confinement, solving this equation is complicated by the following factors: 1) the anisotropy resulting from the large difference in diffusion coefficient in each direction and 2) the instability caused by cross-diffusion terms. The cross-diffusion terms are especially very difficult to handle because they cause failure of the diagonal dominance of the coefficient matrix.

One approach to avoid the cross-diffusion terms is to use a magnetic-field-aligned mesh (MFAM) \cite{Mikellides}. Because the cross-diffusion terms stem from the angle between the magnetic lines of force and the computational mesh, they can be neglected if the mesh is precisely aligned with the magnetic lines of force. However, using an MFAM makes it impossible to use a structured mesh for the body-fitted coordinate system and complicates the evaluation of fluxes on the boundaries. Furthermore, once the magnetic field induced by the plasma flow is solved, one needs to reconstruct the mesh with changing magnetic lines of force. Thus, a practical application of an MFAM is associated with cumbersome steps in coding and implementation.

Recently, a new approach to solving diffusion equations that utilizes a hyperbolic-equation system (HES) has been proposed \cite{Nishikawa2007315,Nishikawa20103989}. This approach divides the diffusion equation into an HES that contains only advection terms by introducing some pseudo-time advancement terms. The mathematical strategy of this approach entails dividing a second or higher order differential equation into a set of first-order differential equations by introducing new variables, including a first derivative of another variable. This approach is supposed to be effective in solving the anisotropic diffusion equation of electron fluids including cross-diffusion terms.

The purpose of this paper is to find a robust and efficient method using an HES for electron fluids in quasi-neutral plasmas. The fundamental equations of electron fluids in quasi-neutral plasmas are introduced and the HES construction for the equations of electron fluids is discussed. Numerical computational fluid dynamics techniques are applied to the HES approach to reduce the computation cost. The HES approach is validated through a two-dimensional numerical test. Also, the performance of the HES approach is analyzed by comparing the computation time of the HES approach with that of the approach using an EE with an MFAM.

\section{Modeling of electron fluids in quasi-neutral plasmas}
\subsection{Fundamental equations}
The fundamental equations treated herein are the two-dimensional conservation equations of electron mass and electron momentum in a quasi-neutral plasma. To solve for the space potential, one generally assumes the plasma approximation, in which the space potential is calculated through conservation equations of electrons, rather than through Gauss's law~\cite{chen1984introduction}. The conservation of mass is formulated as follows:
\begin{equation}
   \nabla \cdot \left(m_{\rm e}n_{\rm e} \vec{u}_{\rm e}\right)= m_{\rm e}n_{\rm e}\nu_{\rm {ion}},
   \label{eq:14}
\end{equation}
where \(m_{\rm e}\), \(n_{\rm e}\), \(\vec{u}_{\rm e}\), and \(\nu_{\rm {ion}}\) are the electron mass, electron number density, electron velocity, and ionization collision frequency, respectively. Under the quasi-neutrality assumption, the electron number density is equal to the ion number density. The ion number density is derived through the computation of the ion flow, which is not treated in this paper.
Thus, the electron number density is treated as a given distribution. The time-derivative term of the electron number density is not included in Eq. (\ref{eq:14}) because electrons are sufficiently mobile to instantaneously achieve quasi-neutrality. Therefore, the electron number density is regarded as a time-invariant quantity.
Because the electron density is given as a time-invariant quantity, the electron fluids in quasi-neutral plasmas have characteristics similar to those of incompressible fluids.

Conservation of momentum is derived from the Navier-Stokes equation,
\begin{equation}
   m_{\rm e}n_{\rm e}\frac{\partial \vec{u}_{\rm e}}{\partial t}+ m_{\rm e}n_{\rm e}\left(\vec{u}_{\rm e} \cdot \nabla\right)\vec{u}_{\rm e}= -\nabla \left(en_{\rm e}T_{\rm e}\right) + en_{\rm e}\nabla{\phi}-en_{\rm e} \vec{u}_{\rm e}\times \vec{B}-m_{\rm e}n_{\rm e}\nu_{\rm {col}}\vec{u}_{\rm e},
   \label{eq:2}
\end{equation}
where \(e\), \(T_{\rm e}\), \(\phi\), \(\vec{B}\), and \(\nu_{\rm {col}}\) are the elemental charge, electron temperature, space potential, magnetic flux density, and electron-neutral total collision frequency, respectively. The forces working on the fluid element are pressure, the electrostatic force, the electromagnetic force, and the collisional force from the collisions between electrons and neutral particles. In this paper, electron temperature is supposed to be a given spatial distribution and to be a time-invariant quantity. This assumption is often used in the simulations of space plasma \cite{Kajimura:2012aa}. If one needs to treat the electron temperature as a time-dependent variable, the electron temperature is calculated by including the energy conservation equation into the system \cite{69922}. In this paper, the energy conservation equation is excluded from the system to focus on a simple anisotropic diffusion equation. Here, the effects of electron-electron collision and electron-ion collision are neglected because generally their collision frequencies are much lower than the electron-neutral collision frequency. Furthermore, because of the large number of collisions, the inertia of the electron fluid is negligibly small, and the force working on the fluid element is balanced, so
\begin{equation}
   0= -\nabla \left(en_{\rm e}T_{\rm e}\right) + en_{\rm e}\nabla{\phi}-en_{\rm e} \vec{u}_{\rm e}\times \vec{B}-m_{\rm e}n_{\rm e}\nu_{\rm {col}}\vec{u}_{\rm e}.
   \label{eq:16}
\end{equation}
After linear conversion of Eq. (\ref{eq:16}), the electron flux in tangential (\(||\)) and orthogonal (\(\perp\)) directions of the magnetic lines of force can be described by using the electron mobility \(\mu\):
\begin{equation}
   n_{\rm e}\left(\begin{array}{c}u_{||} \\ u_{\perp} \end{array}\right)= n_{\rm e}\left[\mu \right]_{\rm {mag}}
   \left(\begin{array}{c}\nabla_{||}\phi \\ \nabla_{\perp}\phi \end{array}\right)-\left[\mu\right]_{\rm {mag}}
   \left(\begin{array}{c}\nabla_{||}\left(n_{\rm e}T_{\rm e}\right) \\ \nabla_{\perp}\left(n_{\rm e}T_{\rm e}\right) \end{array}\right),
   \label{eq:17}
\end{equation}

\begin{equation}
   \left[\mu\right]_{\rm {mag}}= \left[
\begin{array}{cccc}
\mu_{||}  & \    \\
\  & \mu_{\perp}  
\end{array}
\right]= \left[
\begin{array}{cccc}
\frac{e}{m_{\rm e}\nu_{\rm {col}}}  & \    \\
\  & \frac{\mu_{||}}{1+\left(\mu_{||}\left|B \right|\right)^2}
\end{array}
\right].
	\label{eq:18}
\end{equation}
Eq. (\ref{eq:17}) is Ohm's law for electron current on a coordinate system fitted to magnetic lines of force. In the derivation of Eq. (\ref{eq:17}), the E \(\times\) B drift and the diamagnetic drift are neglected by assuming symmetry in one orthogonal direction with the magnetic lines of force. \(\mu_{\perp}\) in Eq. (\ref{eq:18}) is based on the classical diffusion model, and this model can be modified for better reflection of magnetic confinements such as the Bohm diffusion model, depending on the situation \cite{Koo}. The electron mobility on a computational mesh is derived by rotating \(\left[\mu\right]_{\rm {mag}}\) with the angle between the magnetic lines of force and the computational mesh. Thus, we have the following equations.
\begin{equation}
   n_{\rm e}\vec{u_{\rm e}}= n_{\rm e}\left[\mu \right]\nabla \phi-\left[\mu\right]\nabla\left(n_{\rm e}T_{\rm e}\right),
   \label{eq:19}
\end{equation}
\begin{equation}
   \left[\mu\right]=\left[
\begin{array}{cccc}
\mu_{\rm x}  & \mu_{\rm c}    \\
\mu_{\rm c}  & \mu_{\rm y}  
\end{array}
\right]=\Theta^{-1}\left[\mu\right]_{\rm {mag}}\Theta,
\end{equation}
\begin{equation}
\Theta=\left[
\begin{array}{cccc}
 \cos\theta  & -\sin\theta    \\
 \sin\theta  & \cos\theta   
\end{array}
\right].
\end{equation}
Here \(\Theta\) is the rotation matrix and \(\theta\) is the angle between the magnetic lines of force and the computational mesh. 

\subsection{Review of conventional approaches using an elliptic equation}
A common approach to computing the space potential is to utilize an integrated form of the mass conservation equation and momentum conservation equations \cite{Komurasaki:1995fk,Rognlien1992347}. Substituting the divergence of Eq. (\ref{eq:19}) into Eq. (\ref{eq:14}) gives
\begin{equation}
   \nabla\cdot\left(n_{\rm e}\left[\mu \right]\nabla \phi-\left[\mu\right]\nabla\left(n_{\rm e}T_{\rm e}\right)\right)=n_{\rm e}\nu_{\rm {ion}}.
  	\label{eq:21}
\end{equation}
This equation is solved for the space potential as a boundary value problem. This approach is similar to the marker-and-cell (MAC) approach for incompressible fluids in the sense that both utilize an elliptic equation derived by the divergence of the momentum conservation equation \cite{harlow1965numerical}. The approach using Eq. (\ref{eq:21}) is referred to as the EE approach in this paper.

In the electron mobility tensor in Eq. (\ref{eq:21}), \(\mu_{\rm x}\) and \(\mu_{\rm y}\) are coefficients for diffusion terms of \(\partial^2 /\partial x^2\) and \(\partial^2 /\partial y^2\), and \(\mu_{\rm c}\) is the coefficient for the cross-diffusion terms of \(\partial^2 /\partial x\partial y\). Both these two types of diffusion terms pose difficulties if Eq. (\ref{eq:21}) is solved with an iterative method. The first difficulty is due to the large difference between \(\mu_{x}\) and \(\mu_{y}\), which increases the condition number of the problem and degrades convergence performance.

\begin{figure}[h]
	\begin{center}
		\includegraphics[width=120mm]{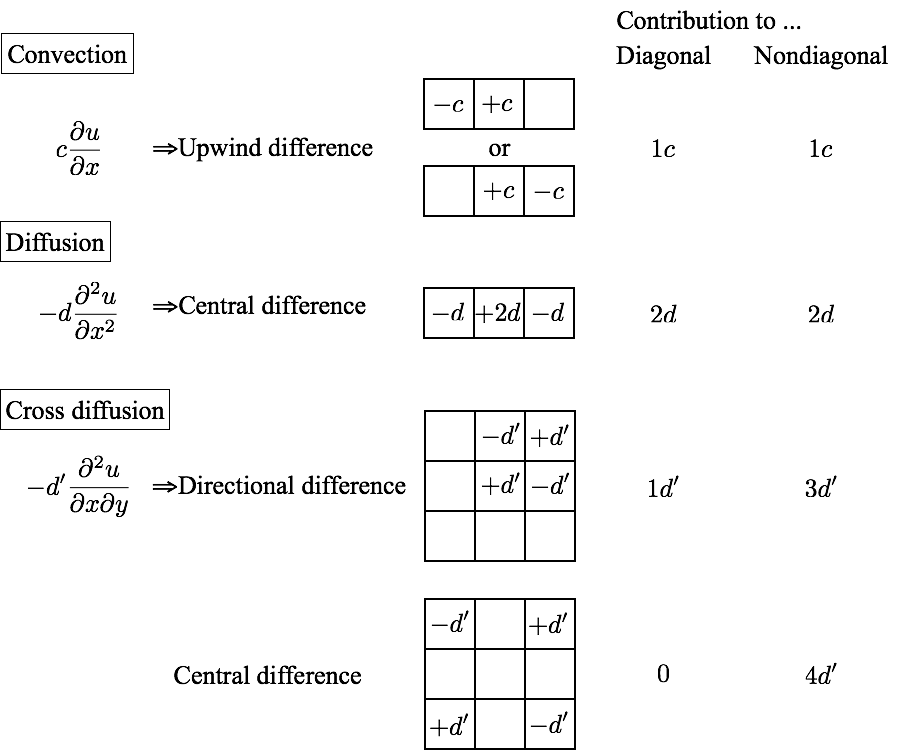}
	\end{center}
	\caption{Contribution of coefficients in stencils of convection, diffusion, and cross-diffusion terms to diagonal and nondiagonal elements of the coefficient matrix. The balance of diagonal and nondiagonal terms is violated in differencing cross-diffusion terms.}
	\label{fig:diagonal}
\end{figure}

The more critical difficulty is that the cross-diffusion terms violate the diagonal dominance of the coefficient matrix. 
The contribution of the coefficient in stencils to diagonal and nondiagonal elements of the coefficient matrix in one row is shown in Fig. \ref{fig:diagonal} for convection, diffusion, and cross-diffusion terms. Convection and diffusion terms increase both the diagonal and nondiagonal elements by the same value, and diagonal dominance is maintained. However, in differencing cross-diffusion terms, the increase of nondiagonal elements is greater than that of diagonal elements with any discretization method based on linear Taylor expansions. The violation of diagonal dominance of the coefficient matrix constrains the convergence speed by \(\mu_{c}/\Delta x\Delta y\). This restriction is very severe in many cases, which results in quite slow convergence. Although one can use a direct method to compute Eq. (\ref{eq:21}), the computation cost is \(O\left(N_{\rm {cell}}^3\right)\), where \(N_{\rm {cell}}\) is the number of cells. Thus, using a direct method is not desired for practical simulations.

Because the cross-diffusion terms are caused by the angle between the computational mesh and the magnetic lines of force, one effective approach to avoid cross-diffusion terms is to use an MFAM \cite{Mikellides}. By aligning the computational mesh with the magnetic lines of force precisely, the effect of cross diffusion is neglected. If only non-cross-diffusion terms are included in Eq. (\ref{eq:21}), the diagonal dominance of the coefficient matrix is satisfied by applying a central difference to the diffusion terms. However, if the magnetic lines of force near the boundary are neither parallel nor perpendicular to the boundary, the MFAM ceases to be a body-fitted mesh and the estimation of fluxes flowing into the boundary becomes complicated. Also, if the magnetic field is time variant like the induced magnetic field, the mesh needs to be reconstructed every iteration, which results in large computation costs. Using a MFAM is effective in avoiding cross-diffusion terms but its application range is limited.

\section{The hyperbolic-equation system approach}
\subsection{Construction of a hyperbolic-equation system}

Another approach to avoid cross-diffusion terms is the HES approach. In this approach one converts the second-order differential equation into a first-order system by introducing new variables, which include the gradient of another variable. This kind of approach was first proposed by Nishikawa for future application to Navier-Stokes equations \cite{Nishikawa2007315,Nishikawa20103989}. 
Although Nishikawa did not mention cross-diffusion terms in his papers, the HES approach is supposed to be beneficial in solving anisotropic diffusion equations including cross-diffusion terms.

Instead of the elliptic equation of Eq. (\ref{eq:21}), the original mass and momentum conservation equations can be used for an HES. To solve for the space potential in an HES, a pseudo-time advancement term of the space potential is introduced into Eq. (\ref{eq:14}). Also, pseudo-time advancement terms of electron momentum are added to Eq. (\ref{eq:19}) for \(x\) and \(y\) directions. Then the HES of electron fluids in quasi-neutral plasmas can be derived as follows:
\begin{equation}
   \frac{1}{a}\frac{n_{\rm e}}{T_{\rm e}}\frac{\partial \phi}{\partial t}-\nabla \cdot \left(n_{\rm e} \vec{u}_{\rm e}\right)=-n_{\rm e}\nu_{\rm {ion}},
   \label{eq:22}
\end{equation}
\begin{equation}
   \frac{1}{\nu_{\rm {col}}}\left[
\begin{array}{cc}
b_{\rm x} & \   \\
\  & b_{\rm y}
\end{array}
\right]^{-1}
\frac{\partial}{\partial t}\left(n_{\rm e}\vec{u}_{\rm e}\right)
- n_{\rm e}\left[\mu \right]\nabla \phi+\left[\mu\right]\nabla\left(n_{\rm e}T_{\rm e}\right)=-n_{\rm e}\vec{u}_{\rm e},
	\label{eq:23}
\end{equation}
where \(a\), \(b_{\rm x}\), and \(b_{\rm y}\) are arbitrary acceleration parameters of no dimension. 
Because the pseudo-time advancement terms are artificially added, the HES of Eqs. (\ref{eq:22}) and (\ref{eq:23}) needs to be calculated until a steady state is obtained to make the artificial terms negligibly small. In a steady state, the HES is equivalent to the mass and momentum conservation equations. This approach is similar to the pseudo-compressibility approach for incompressible fluids in the sense that both introduce a pseudo-time advancement term into the conservation equation \cite{Chorin196712}. However, the key point of this approach is to add the pseudo-time advancement term of the space potential in the mass conservation equation, but not that of electron density.

\subsection{Nondimensional form of the hyperbolic-equation system}
Eqs. (\ref{eq:22}) and (\ref{eq:23}) are modified to a nondimensional form for the analysis. First, the electron mobility tensor is normalized by the electron mobility in the tangential direction of the magnetic lines of force:
\begin{equation}
   \tilde{\left[\mu\right]}=\left[
\begin{array}{cccc}
\tilde{\mu}_{\rm x}  & \tilde{\mu}_{\rm c}    \\
\tilde{\mu}_{\rm c}  & \tilde{\mu}_{\rm y}  
\end{array}
\right]=\frac{1}{\mu_{||}}\left[\mu\right]
=\frac{m_{\rm e}\nu_{\rm {col}}}{e}\left[\mu\right].
\end{equation}
By using representative values of electron number density \(n_{\rm e}^*\), electron temperature \({T_{\rm e}^*}\) and mean free path \(\lambda_{\rm m}^*\), the nondimensional values of the physical quantities are defined as follows:
\begin{equation}
   \tilde{n}_{\rm e}=\frac{n_{\rm e}}{n_{\rm e}^*},\hspace{20pt}
   \tilde{T}_{\rm e}=\frac{T_{\rm e}}{T_{\rm e}^*},\hspace{20pt}
   \tilde{\phi}=\frac{\phi}{T_{\rm e}^*},\hspace{20pt}
   \left(\tilde{x}, \tilde{y}\right)^T=\frac{1}{\lambda_{\rm m}^*}\left(x, y\right)^T,
\end{equation}
\begin{equation}
   \tilde{t}=\frac{1}{\tau_{\rm m}^*}t=\frac{v_{\rm {e,th}}^*}{\lambda_{\rm m}^*}t=\frac{1}{\lambda_{\rm m}^*}\sqrt{\frac{2eT_{\rm e}^*}{m_{\rm e}}}t,\hspace{20pt}
   \tilde{\vec{u}}_{\rm e}=\frac{\vec{u}_{\rm e}}{c_{\rm s}^*}=\frac{\vec{u}_{\rm e}}{\sqrt{\frac{\gamma eT_{\rm e}^*}{m_{\rm e}}}},\hspace{20pt}   
   \tilde{\nu}_{\rm {col}}=\tau_{\rm m}^*\nu_{\rm {col}},\hspace{20pt}
   \tilde{\nu}_{\rm {ion}}=\tau_{\rm m}^*\nu_{\rm {ion}}.
\end{equation}
Here the tilde denotes a nondimensional quantity. \(\tau_{\rm m}\), \(v_{\rm {e,th}}\), and \(c_{\rm s}\) are the mean free time, electron thermal velocity, and electron acoustic velocity, respectively. By using these quantities, a nondimensional equation system can be constructed. Furthermore, \(\tilde{n}_{\rm e}=1\), \(\tilde{T}_{\rm e}=1\), \(\tilde{\nu}_{{\rm col}}=1\), and \(\tilde{\nu}_{\rm {ion}}=0\) are assumed throughout the calculation region for simplified analysis of the space potential and electron velocity. Eventually, the simplified nondimensional equation system can be expressed as follows:
\begin{equation}
   \frac{1}{a}\frac{\partial \tilde{\phi}}{\partial \tilde{t}}-\sqrt{\frac{\gamma}{2}}\tilde{\nabla} \cdot \tilde{\vec{u}_{\rm e}}=0,
   \label{eq:22-1}
\end{equation}
\begin{equation}
\left[
\begin{array}{cc}
b_{\rm x} & \   \\
\  & b_{\rm y}
\end{array}
\right]^{-1}
\frac{\partial \tilde{\vec{u}_{\rm e}}}{\partial \tilde{t}}
- \frac{1}{\sqrt{2\gamma}}\left[\tilde{\mu} \right]\tilde{\nabla} \tilde{\phi}=-\tilde{\vec{u}_{\rm e}}.
	\label{eq:23-1}
\end{equation} 

Optimal choice of the acceleration parameters improves the condition of the numerical problem. One of the criteria controlling the difficulty of the numerical problem is the condition number. In this problem the condition number can be interpreted as the absolute value of the ratio of the maximum eigenvalue to the minimum eigenvalue in the coefficient matrix. Thus, to make the eigenvalues unity in each direction the acceleration coefficients are chosen as follows:
\begin{equation}
   a=\sqrt{\frac{2}{\gamma}},\hspace{40pt}
   b_{\rm x}=\frac{\sqrt{2\gamma}}{\tilde{\mu}_{\rm x}},\hspace{40pt}
   b_{\rm y}=\frac{\sqrt{2\gamma}}{\tilde{\mu}_{\rm y}}.
\end{equation}
With these acceleration coefficients, Eqs. (\ref{eq:22-1}) and (\ref{eq:23-1}) can be rewritten in the vector form:
\begin{equation}
   \frac{\partial U}{\partial \tilde{t}}+J_x\frac{\partial U}{\partial \tilde{x}}+J_y\frac{\partial U}{\partial \tilde{y}}=S,
   \label{eq:vector}
\end{equation}
where
\begin{equation}
   U=\left(\tilde{\phi},\tilde{u}_{\rm x},\tilde{u}_{\rm y}\right)^T,\hspace{20pt}
   S=\left(0,-\frac{\sqrt{2\gamma}}{\tilde{\mu}_{\rm x}}\tilde{u_{\rm x}},-\frac{\sqrt{2\gamma}}{\tilde{\mu}_{\rm y}}\tilde{u_{\rm y}}\right)^T
   \label{eq:variable}
\end{equation}
\begin{equation}
   J_{x}=\left(
\begin{array}{ccc}
0 &-1 & 0   \\
-1 & 0 & 0 \\
-\frac{\tilde{\mu}_{\rm c}}{\tilde{\mu}_{\rm y}} & 0 & 0 
\end{array}
\right),\hspace{20pt}
J_{y}=\left(
\begin{array}{ccc}
0 & 0 &-1   \\
-\frac{\tilde{\mu}_{\rm c}}{\tilde{\mu}_{\rm x}} & 0 & 0 \\
-1 & 0 & 0 
\end{array}
\right).
	\label{eq:25}
\end{equation}
Here, \(J_x\) and \(J_y\) are the Jacobian matrices in the \(x\) and \(y\) directions. The eigenvalues for the Jacobian matrices are as simple as follows:
\begin{equation}
   \lambda_{x}=0, \pm 1,
	\hspace{20pt}
	\lambda_{y}=0, \pm 1.
	\label{eq:26}
\end{equation}
By using the Jacobian matrices and the eigenvalues, upwind schemes based on the approximate Riemann solver can be constructed.
\section{Test calculation condition and numerical method}

\subsection{Calculation condition in two dimensions}
\begin{figure}[t]
  \begin{center}
    \includegraphics[width=100mm]{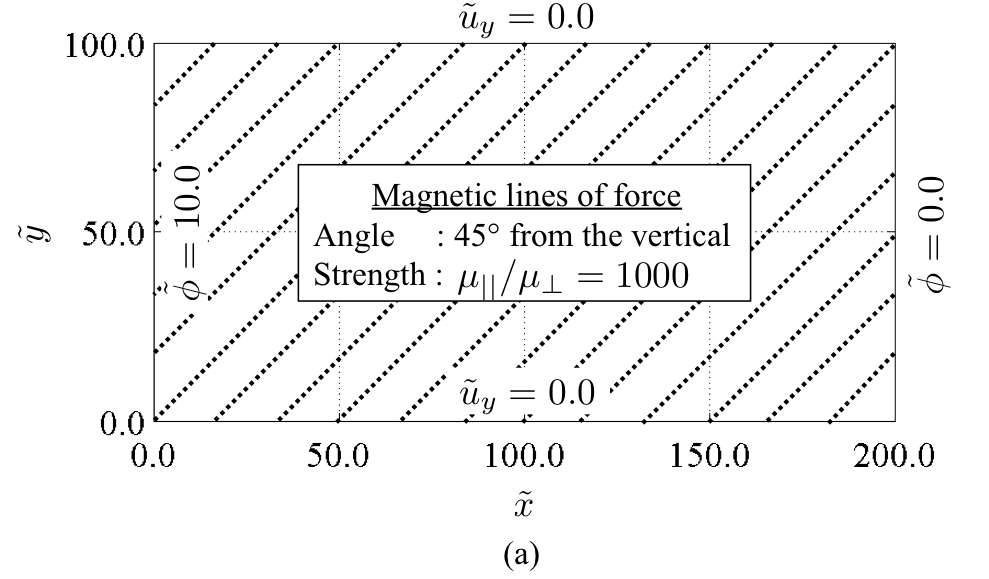}
  \end{center}
  \begin{center}
    \includegraphics[width=100mm]{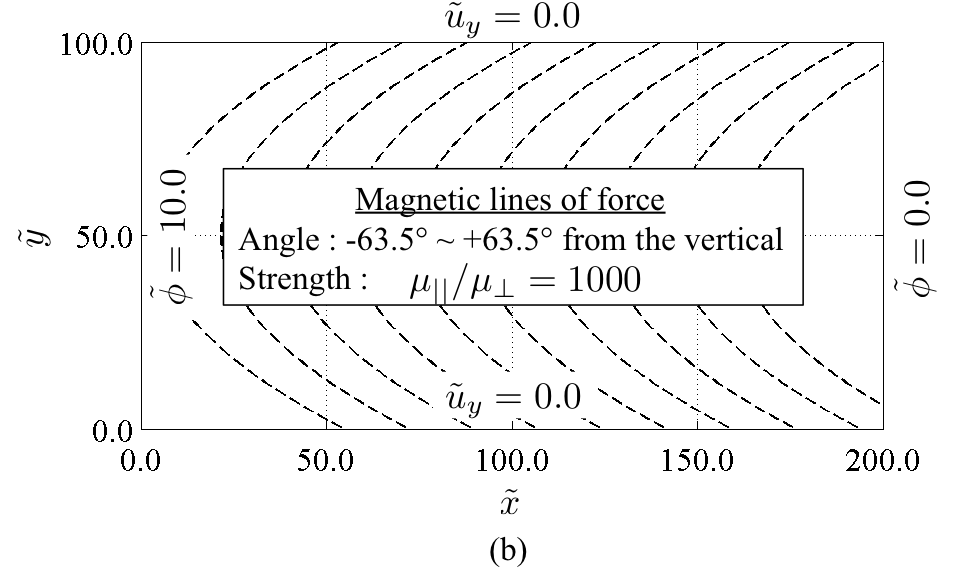}
  \end{center}
  \caption{Two-dimensional calculation conditions used for the test calculation. (a): The magnetic lines of force are uniformly angled by 45\(^\circ\). (b): A concave shape of magnetic lines of force toward right-hand side are applied with the angle from the vertical of -63.5\(^\circ\) at the bottom and 63.5\(^\circ\) at the top of the calculation field. For both cases the strength of magnetic confinement of \(\mu_{||}/\mu_{\perp}=1000\) is assumed. Dirichlet conditions are used on \(\tilde{\phi}\) for the left and right side boundaries. The zero-flux condition is used for the top and bottom boundaries.}
	\label{fig:2}
\end{figure}

Test calculations in two dimensions are conducted for the analyses of the HES approach for the electron fluid equations. Another intention of the test is to compare the HES and EE approaches in terms of the computation cost. The calculation condition is illustrated in Fig. \ref{fig:2} (a). Magnetic lines of force, uniformly angled at 45\(^\circ\) from the vertical and with uniform strength of magnetic confinement of \(\mu_{||}/\mu_{\perp}\) = 1000, are applied on the calculation field. Dirichlet conditions on the nondimensional space potential are defined at the left and right side boundaries. Zero-flux conditions are used for the top and bottom boundaries. The condition of 45\(^\circ\) magnetic lines of force gives a maximum effect of cross diffusion in the EE approach when a vertical-horizontal mesh (VHM) is used. Thus if this condition can be solved stably with the HES approach, robust calculations for any angle of magnetic lines of force can be expected. 

Also, to confirm that the HES approach is applicable to a condition with complicated configuration of magnetic lines of force, a condition of concave magnetic lines of force is calculated with the HES approach. This calculation condition is illustrated in Fig. \ref{fig:2} (b). The magnetic lines of force are angled by -63.5\(^\circ\) to 63.5\(^\circ\) from the vertical line in the calculation field. The uniform strength of magnetic confinement
of \(\mu_{||}/\mu_{\perp}\) = 1000 is assumed. Nondimensional space potential is defined at the left and right boundaries, and nondimensional electron velocity in y-direction is defined for the top and bottom boundaries.

\subsection{Numerical method used in the HES approach}
The numerical methods used for the HES approach are summarized in Table \ref{t1}. The discretization is implemented based on a finite-difference method. A first-order upwind scheme is constructed based on the Jacobian matrices and the eigenvalues of Eqs. (\ref{eq:25}) and (\ref{eq:26}). Here the upwind difference based on an approximate Riemann solver of Steger-Warming's flux vector splitting (FVS) is used \cite{Steger1981263}. The FVS is modified for the non-conservation form of Eq. (\ref{eq:vector}) \cite{Gabutti1983207}. For instance, the space difference in the \(\tilde{x}\) direction is split as follow:
\begin{equation}
	J_{\rm x}\frac{\partial U}{\partial \tilde{x}}\simeq\frac{1}{\Delta \tilde{x}}J_{\rm x}^+\delta_{\rm x}^b U+\frac{1}{\Delta \tilde{x}}J_{\rm x}^-\delta_{\rm x}^f U.
	\label{eq:fvs}
\end{equation}
Here, \(J_{\rm x}^+\) and \(J_{\rm x}^-\) are the Jacobian matrices which have only positive and negative eigenvalues, and \(\delta_{\rm x}^f\) and \(\delta_{\rm x}^b\) are the forward and backward derivatives, respectively.
Because the signs of the eigenvalues are fixed, the split terms in Eq. (\ref{eq:fvs}) are always differentiable. Thus the scheme does not lead to instability associated with the change of signs in eigenvalues. 

Implicit methods should be used for the pseudo-time advancement method because only the steady-state solution is needed. In an implicit method, the space difference is evaluated at the \(n+1\) time level. Eq. (\ref{eq:vector}) can be written by discretizing the time-derivative term in the first-order form as:
\begin{equation}
	\frac{\Delta U^n}{\Delta \tilde{t}}+\left(J_x\frac{\partial U}{\partial \tilde{x}}+J_y\frac{\partial U}{\partial \tilde{y}}\right)^{n+1}=S^{n},
\end{equation}
where \(\Delta \tilde{t}\) is the time step and \(\Delta U^n=U^{n+1}-U^n\).
This formulation is converted to a ``delta" form of the implicit method by utilizing Beam-Warming linearization \cite{Beam:1978fj}. 
\begin{equation}
	\left(I+\Delta\tilde{t}J_x^n\frac{\partial}{\partial \tilde{x}}+\Delta\tilde{t}J_y^n\frac{\partial}{\partial \tilde{y}}\right){\Delta U^n}=-\Delta\tilde{t}\left(J_x\frac{\partial U}{\partial \tilde{x}}+J_y\frac{\partial U}{\partial \tilde{y}}\right)^n+\Delta\tilde{t}S^{n}.
	\label{eq:implicit}
\end{equation}

For the two-dimensional calculation, the alternating direction implicit (ADI) method is used for fast computation. In the EE approach, the ADI method cannot be used if there are cross-diffusion terms. Information on the stencils of \(\left(i+1, j+1\right)\), \(\left(i-1, j+1\right)\), \(\left(i+1, j-1\right)\), or \(\left(i-1, j-1\right)\) must be used for cross-diffusion terms if the differencing is based on a linear Taylor series in two dimensions. The coefficients for these stencils appear in isolated positions from diagonal positions in the coefficient matrix; thus reduction of computation cost by ADI is not expected. However, in the HES approach, the ADI method can be used because cross-diffusion terms are not included. The ADI factorization proposed by Beam and Warming is applied to Eq. (\ref{eq:implicit}) as follow \cite{Beam:1978fj}:
\begin{equation}
	\left(I+\Delta\tilde{t}J_x^n\frac{\partial}{\partial \tilde{x}}\right)\left(I+\Delta\tilde{t}J_y^n\frac{\partial}{\partial \tilde{y}}\right){\Delta U^n}=-\Delta\tilde{t}\left(J_x\frac{\partial U}{\partial \tilde{x}}+J_y\frac{\partial U}{\partial \tilde{y}}\right)^n+\Delta\tilde{t}S^{n}.
	\label{eq:adi}
\end{equation}
To strengthen the diagonal dominance, an lower-upper ADI (LU-ADI) scheme is further implemented \cite{FUJII:1986xy}. In the LU-ADI method, for instance, the operator in the \(\tilde{x}\) direction can be written as follow:
\begin{equation}
	I+\Delta\tilde{t}J_x\frac{\partial}{\partial \tilde{x}} \simeq 
	\left(I-\frac{\Delta\tilde{t}}{\Delta\tilde{x}}J_x^-+\Delta\tilde{t}J_x^+\delta_x^b\right)
	\left(I+\frac{\Delta\tilde{t}}{\Delta\tilde{x}}\left(J_x^+-J_x^-\right)\right)^{-1}
	\left(I+\frac{\Delta\tilde{t}}{\Delta\tilde{x}}J_x^++\Delta\tilde{t}J_x^-\delta_x^f\right).
	\label{eq:luadi}
\end{equation}
The computations are implemented with a Courant number of 30. 

For the HES approach, a VHM is used as shown in Fig. \ref{fig:3} (a). The boundary conditions are set at the ``ghost cells" outside of the calculation field. According to the eigenvalues of Eq. (\ref{eq:26}), one characteristic speed is flowing from outside to inside of the boundary. Thus, among the three variables of Eq. (\ref{eq:variable}), one variable should be defined as the boundary condition. The calculation condition of Fig. \ref{fig:2} satisfies this requirement on the boundary conditions.

\subsection{Numerical method used in the EE approach}
\begin{figure}[t]
  \begin{minipage}{0.5\hsize}
  \begin{center}
      \includegraphics[width=60mm]{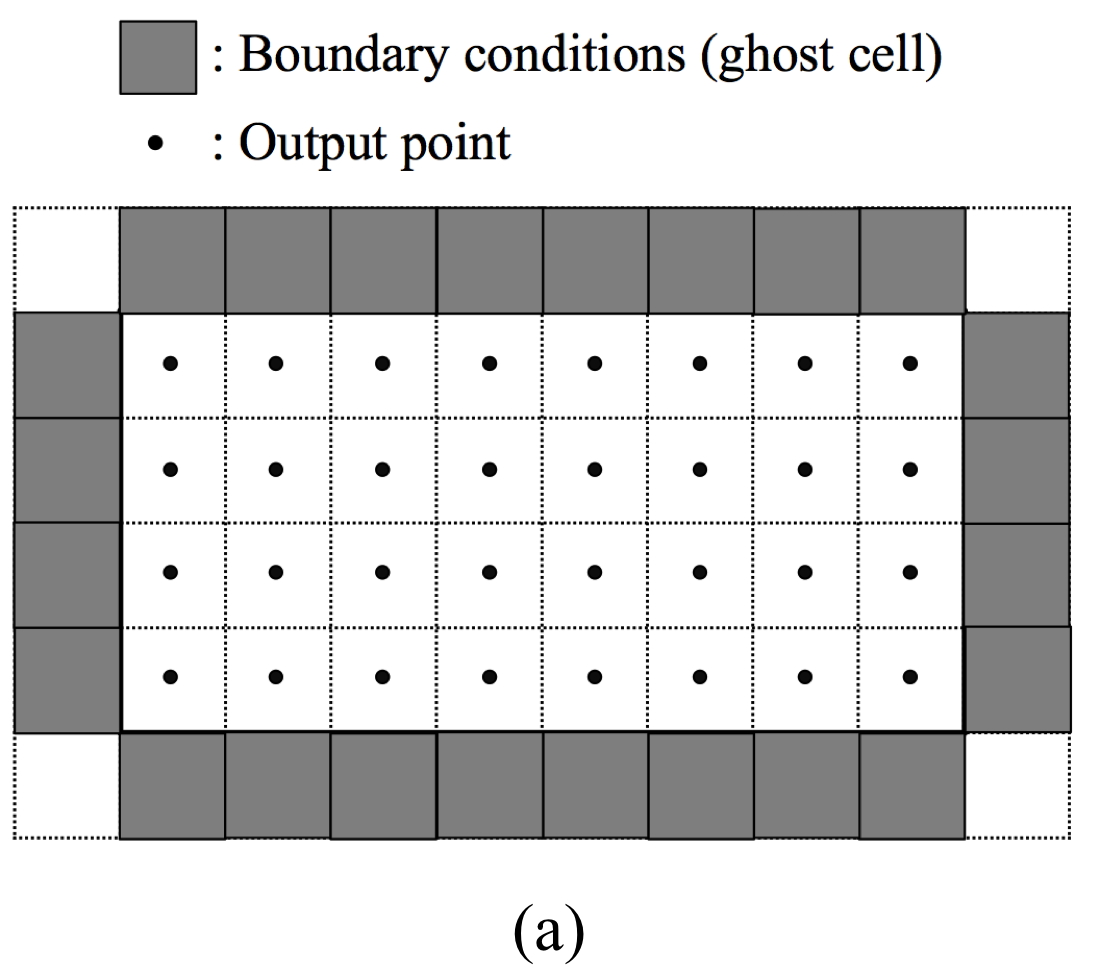}
  \end{center}
 \end{minipage}
 \begin{minipage}{0.5\hsize}
  \begin{center}
      \includegraphics[width=60mm]{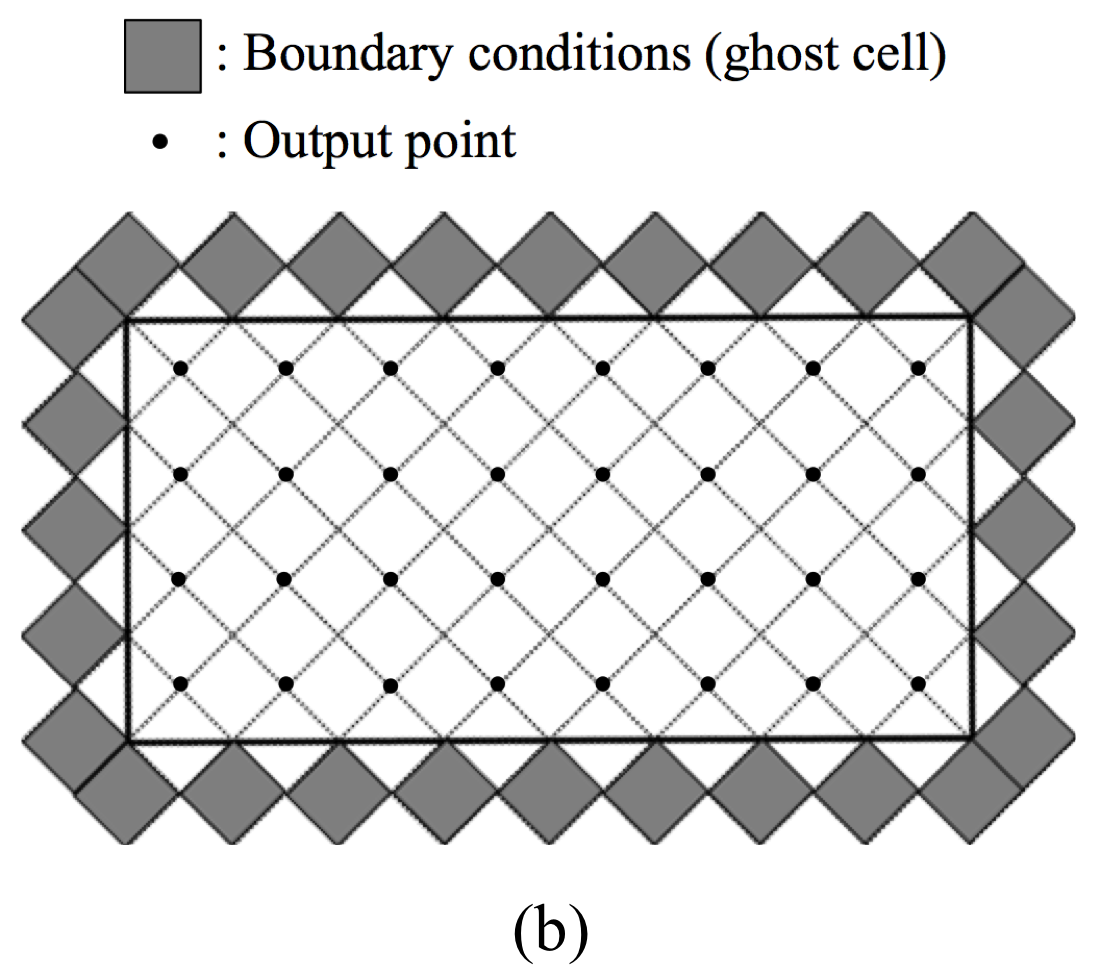}
  \end{center}
 \end{minipage}
  \caption{(a): A VHM for a grid of 8 \(\times\) 4 (output points). (b): An MFAM for a grid of 8 \(\times\) 4 (output points).}
	\label{fig:3}
\end{figure}

The numerical methods used for the EE approach are also summarized in Table \ref{t1}. The EE approach with an MFAM handles only non-cross-diffusion terms. Thus a second-order central difference is used with the finite-volume method. To solve the boundary value problem of the elliptic equation, a successive over-relaxation (SOR) method is used. In the test calculations the relaxation parameter of the SOR is set as 1.8. For the EE approach, an MFAM is used, as is shown in Fig. \ref{fig:3} (b). This mesh is derived by rotating the VHM by 45\(^\circ\). The boundary conditions are defined at ghost cells outside the calculation field. For the upper and lower boundary conditions of \(u_{\rm y} = 0\), the electron mass flux of Eq. (\ref{eq:21}) is set to zero with a finite-volume method.

\begin{table}
\caption{Summary of the HES and EE approaches.}
\label{t1}
    \begin{center}
\begin{tabular}{lll}
	\Hline
	\  & HES approach & EE approach \\
	\hline
	Equation & Hyperbolic-equation system (HES) & Elliptic equation (EE) \\ 
	Mesh & Vertical-horizontal mesh (VHM) & Magnetic-field-aligned mesh (MFAM) \\
	Scheme & First-order upwind differencing & Second-order central differencing \\
	Iteration & LU-ADI method & Successive over relaxation (SOR) method \\
	\Hline
	\end{tabular}
    \end{center}
\end{table}

\section{Results and discussion}

\subsection{Steady-state calculation results}
\begin{figure}[h]
    \begin{center}
      \includegraphics[width=90mm]{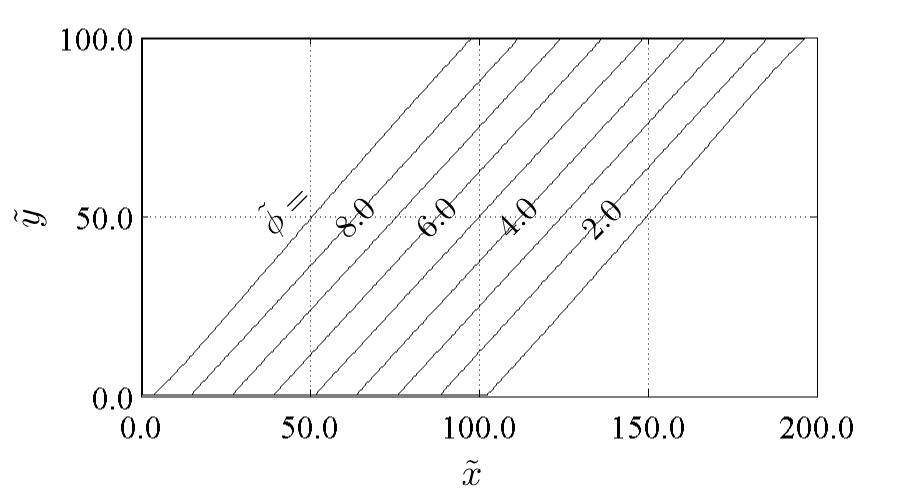}
    \end{center}
    \begin{center}
      \includegraphics[width=90mm]{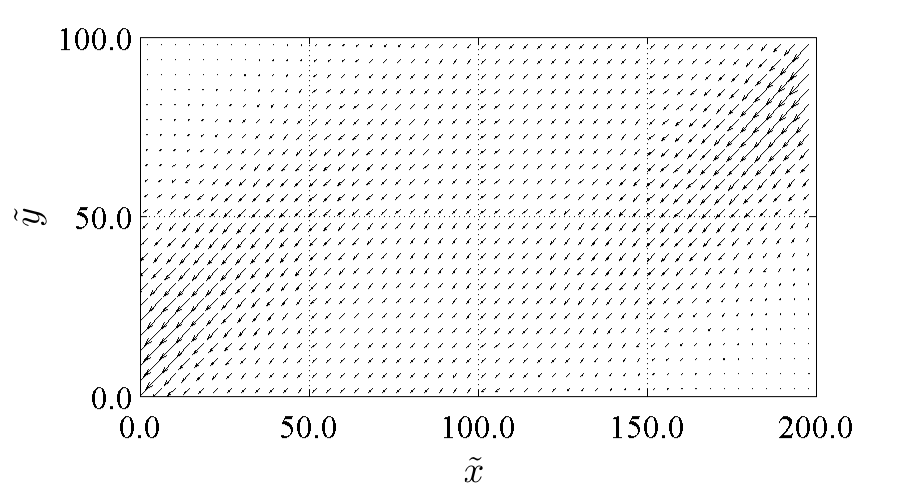}
    \end{center}
  \caption{Simulation results for the Fig. \ref{fig:2} (a) condition with the HES approach using a VHM grid of 48 \(\times\) 24. Top: Nondimensional space potential distribution. Bottom: Vector map of nondimensional velocity.}
  \label{fig:re45_1}
    \begin{center}
      \includegraphics[width=90mm]{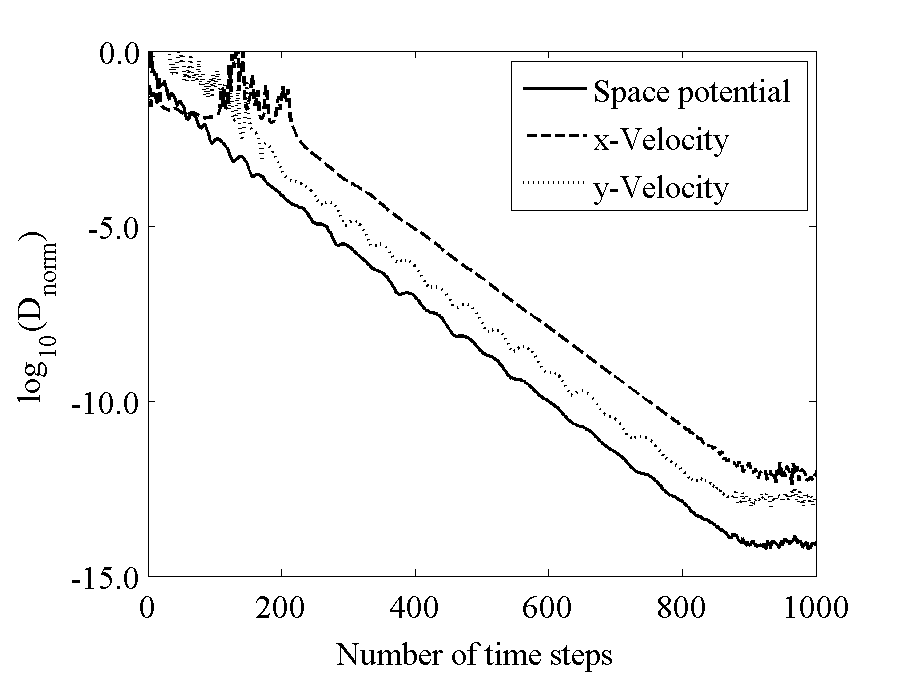}
    \end{center}
  \caption{Time history of the normalized difference of each variable. The Fig. \ref{fig:2} (a) condition is solved with the HES approach using a 48 \(\times\) 24 grid.}
  \label{fig:convergence}\end{figure}

The HES approach is tested for the calculation condition of Fig. \ref{fig:2} (a) with a VHM for a grid of 48 \(\times\) 24. The steady-state calculation results are shown in Fig. \ref{fig:re45_1}. The equipotential lines are almost uniformly angled by 45\(^\circ\), which indicates that the result reflects the effect of magnetic confinement. Also, the velocity vector map reflects the zero-flux boundary condition at the top and bottom boundaries. Because the HES approach utilizes pseudo-time advancement terms, they have to converge to be negligibly small values in the steady state. For evaluating the convergence of each variable, the normalized difference is defined as
\begin{equation}
   D_{\rm {norm}}=\sqrt{\frac{1}{N_{\rm {cell}}}\sum^{N_{\rm {cell}}}\left(\frac{\left|y^{n+1}-y^{n}\right|^2}{\left|y^n\right|^2+\varepsilon}\right)}.
   \label{eq:Dnorm}
\end{equation}
Here \(y\) is a variable and \(\varepsilon\) is a positive value satisfying \(\varepsilon\ll \left|y\right|\) to avoid division by zero. The time history of the normalized difference of each variable is shown in Fig. \ref{fig:convergence}. The normalized difference of each variable shows a monotone decrease, and they become negligibly small in the steady state. This fact verifies the usefulness of pseudo-time advancement terms and that a robust calculation is possible with the HES approach for the conditions of angled magnetic lines of force.

\begin{figure}[t]
    \begin{center}
      \includegraphics[width=90mm]{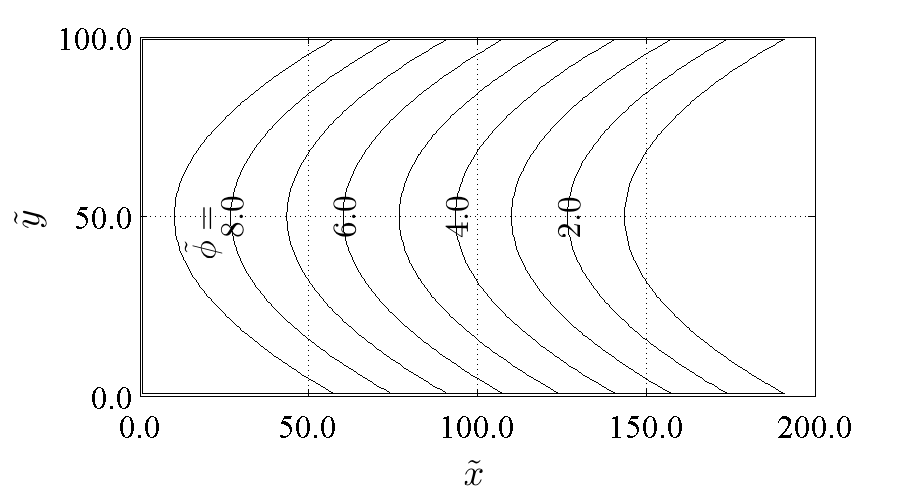}
    \end{center}
    \begin{center}
      \includegraphics[width=90mm]{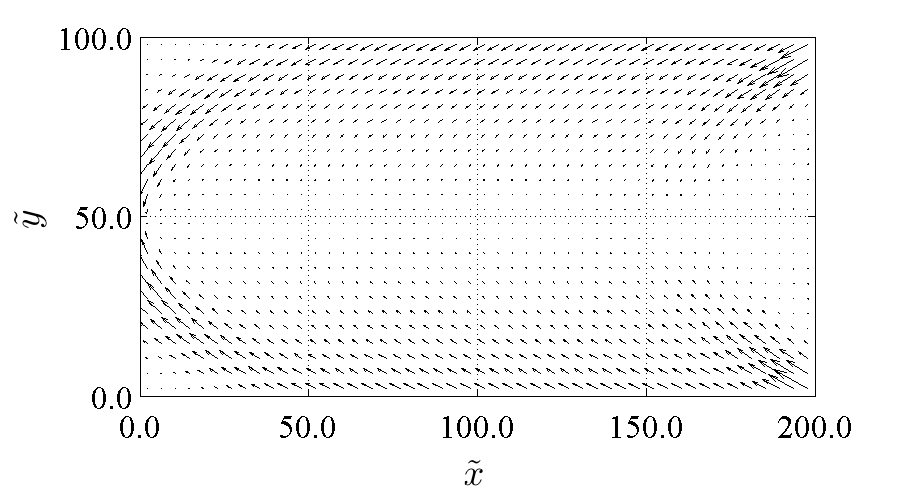}
    \end{center}
  \caption{Simulation results for the Fig. \ref{fig:2} (b) condition with the HES approach using a VHM grid of 48 \(\times\) 24. Top: Nondimensional space potential distribution. Bottom: Vector map of nondimensional velocity.}
  \label{fig:result_curve}

    \begin{center}
      \includegraphics[width=90mm]{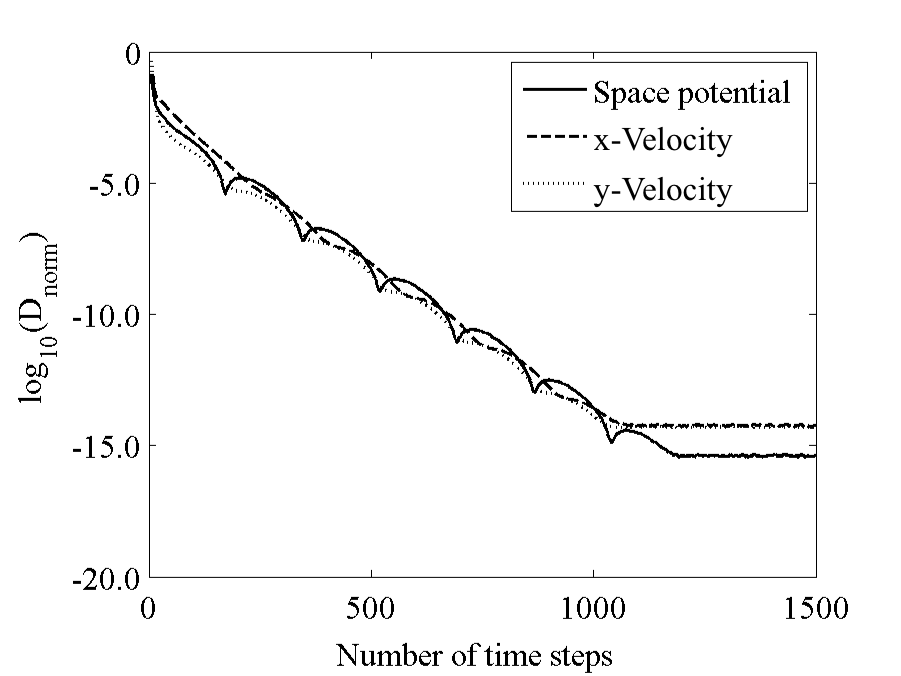}
    \end{center}
  \caption{Time history of the normalized difference of each variable. The Fig. \ref{fig:2} (b) condition is solved with the HES approach using a 48 \(\times\) 24 grid.}
  \label{fig:convergence_curve}
\end{figure}

The condition of Fig. \ref{fig:2} (b) is also calculated by the HES approach using a VHM for a grid of 48 \(\times\) 24. The steady-state calculation results and the time history of the normalized difference are shown in Fig. \ref{fig:result_curve} and Fig. \ref{fig:convergence_curve}, respectively. It is confirmed that the equipotential lines reflect the magnetic lines of force and all of the pseudo-time advancement terms become negligibly small values in the steady state. This indicates that the HES approach can robustly compute the condition of complicated magnetic lines of force configuration.


\clearpage

\subsection{Computation cost comparison}
\label{sec:5.3}
The computation cost of the HES approach is compared with that of the EE approach. Order analyses of computation cost are conducted in terms of the size of the problem and the strength of magnetic confinement. The size of the problem is evaluated with the number of cells, \(N_{\rm {cell}}\), and the strength of magnetic confinement is associated with the ratio of electron mobility, \(\mu_{||}/\mu_{\perp}\). Computation cost is measured with a sequential calculation without parallelization. The convergence is deemed to be satisfied when the normalized difference of the space potential reaches 10\(^{-10}\), and the CPU second at convergence is termed \(T_{\rm {converge}}\).

\begin{figure}[t]
  \begin{center}
    \includegraphics[width=85mm]{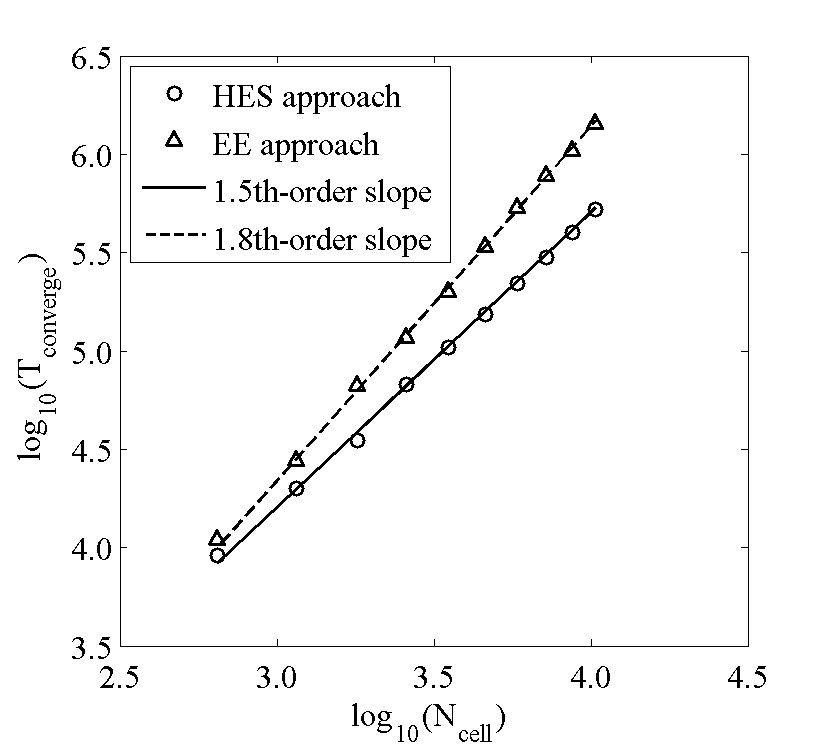}
  \end{center}
\caption{Computation costs until convergence when the number of cells is changed. The Fig. \ref{fig:2} (a) condition is solved using the HES approach and the EE approach. A logarithmic scale is used for each axis for scale analysis. 1.5th-, and 1.8th-order slopes are depicted for reference.}
   \label{fig:cost1}
\end{figure}

Fig. \ref{fig:cost1} shows \(T_{\rm {converge}}\) of the HES approach and the EE approach when the number of cells, \(N_{\rm {cell}}\), is varied. The computation time of the HES approach is \(O\left(N_{\rm {cell}}^{1.5}\right)\), whereas the cost of the EE approach is \(O\left(N_{\rm {cell}}^{1.8}\right)\). From this result, we conclude that the differences in computation time in terms of the size of the problem between the two approaches is insignificant. This fact supports the advantage of the HES approach in its ability to utilize a simple structured mesh without increasing computation time.

Fig. \ref{fig:11} shows a comparison of \(T_{\rm {converge}}\) between the two approaches when the strength of magnetic confinement, \(\mu_{||}/\mu_{\perp}\), is varied. The computation time of the EE approach increases with \(\mu_{||}/\mu_{\perp}\), whereas that of the HES approach stays almost constant. This is because the eigenvalues of the system are always adjusted via Eq. (\ref{eq:26}) by the acceleration coefficient, and the condition number of the system does not increase with \(\mu_{||}/\mu_{\perp}\) in the HES approach. This result indicates that the HES approach is suitable under conditions of strong anisotropy resulting from magnetic confinement. This is also an advantage of the HES approach.

\subsection{Comparison on mesh convergence of transverse electron flux}
\begin{figure}[t]
  \begin{center}
   \includegraphics[width=85mm]{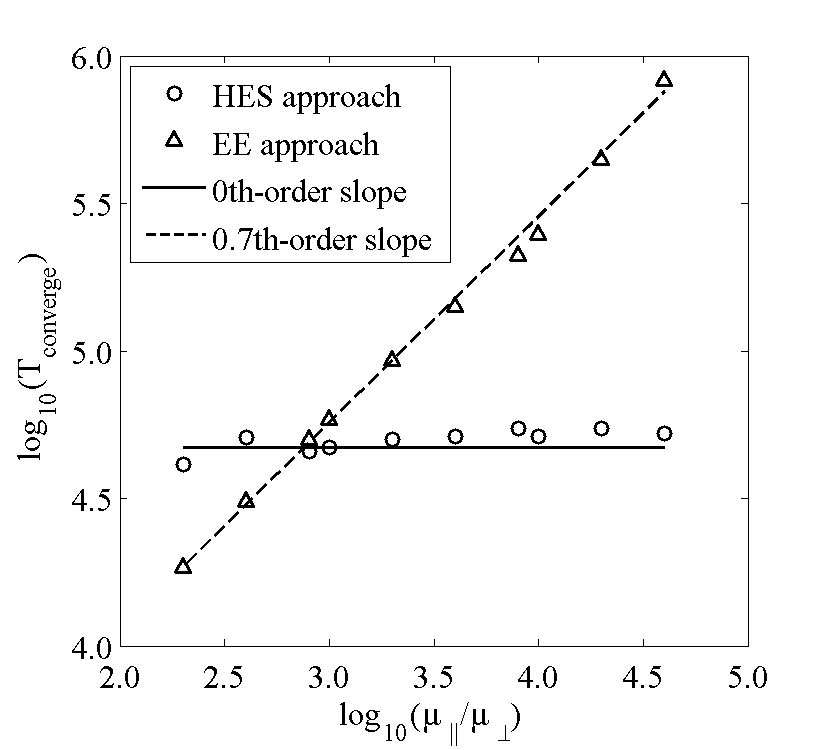}
  \end{center}
\caption{Computation costs until convergence when \(\mu_{||}/\mu_{\perp}\) is changed. The Fig. \ref{fig:2} (a) condition is solved using the HES approach and EE approach. A logarithmic scale is used for each axis for scale analysis. Zeroth-, and 0.7th-order slopes are depicted for reference.}
	\label{fig:11}
  \begin{center}
    \includegraphics[width=85mm]{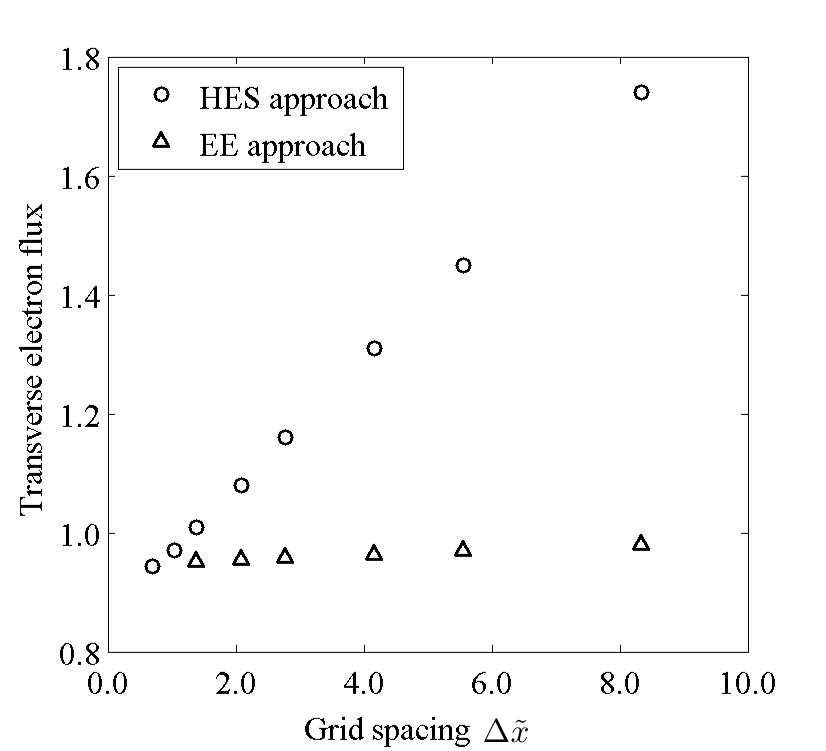}
  \end{center}
\caption{The transverse electron flux calculated by the HES and EE approaches when the grid spacing is varied. The Fig. \ref{fig:2} (a) condition is solved using the HES approach and EE approach. }
   \label{fig:meshconv}
\end{figure}

It is difficult to analyze the computational accuracy for anisotropic diffusion equations because the analytical solution is hard to be derived.
Therefore, the mesh convergence of the calculated electron transverse flux is evaluated for the HES and EE approaches. Here the transverse electron flux is defined as the total electron flux flowing from the right to left side boundaries. The result is visualized in Fig. \ref{fig:meshconv}. 
The transverse electron fluxes calculated with a fine grid system are almost the same between the HES and EE approaches. However, with a coarse grid system, the transverse electron flux is overestimated in the HES approach. This indicates the HES approach has a large numerical viscosity stemming from the first-order upwind scheme. Thus, the HES approach should be used with schemes of high-order spatial accuracy for faster mesh convergence in practical calculations.

\clearpage
\section{Concluding remarks}
A new approach using an HES is proposed for electron fluids in quasi-neutral plasmas. The main advantage of this approach is that it avoids cross-diffusion terms, which violate the diagonal dominance in conventional approaches using the EE. The HES approach enables robust calculations with a simple structured mesh such as a VHM by using an upwind scheme.
Numerical experiments in two dimensions are conducted with angled magnetic lines of force to validate the HES approach. Furthermore, the computation time of the HES approach is compared with that of the conventional EE approach. Also, the mesh convergence of transverse electron flux calculated by the two approaches are compared. The findings are summarized as follows: 

\begin{enumerate}
	\item The calculation results of the HES approach reflect the magnetic confinement and the boundary conditions, and all pseudo-time advancement terms converge to negligibly small values. These facts validate the HES approach.
	\item The computation time of the HES approach is \(O\left(N_{\rm {cell}}^{1.5}\right)\), whereas the cost of the EE approach is \(O\left(N_{\rm {cell}}^{1.8}\right)\). The HES approach can utilize a simple structured mesh without increasing computation time.
	\item With increasing \(\mu_{||}/\mu_{\perp}\), the computation time of the HES approach stays constant, whereas that of the EE approach increases. The HES approach is efficient in solving conditions of strong magnetic confinement.
	\item The HES approach has a large numerical viscosity stemming from the first-order upwind scheme. The HES approach should be implemented with schemes of high-order spatial accuracy for faster mesh convergence.
\end{enumerate}

\section*{Acknowledgment}

This work was supported by the Grant-in-Aid for JSPS Fellows, No. 24-10079.


\renewcommand{\refname}{Reference}
\bibliographystyle{elsarticle-num} 
\bibliography{reference}

\begin{thebibliography}{10}
\expandafter\ifx\csname url\endcsname\relax
  \def\url#1{\texttt{#1}}\fi
\expandafter\ifx\csname urlprefix\endcsname\relax\def\urlprefix{URL }\fi
\expandafter\ifx\csname href\endcsname\relax
  \def\href#1#2{#2} \def\path#1{#1}\fi

\bibitem{Parra}
F.~I. Parra, E.~Ahedo, J.~M. Fife, M.~Mart{\'\i}nez-S{\'a}nchez, A
  two-dimensional hybrid model of the {H}all thruster discharge, Journal of
  Applied Physics 100~(2) (2006) 023304.

\bibitem{kajimura2010hybrid}
Y.~Kajimura, H.~Usui, I.~Funaki, K.~Ueno, M.~Nunami, I.~Shinohara, M.~Nakamura,
  H.~Yamakawa, Hybrid particle-in-cell simulations of magnetic sail in
  laboratory experiment, Journal of Propulsion and Power 26~(1) (2010)
  159--166.

\bibitem{GRL:GRL8860}
D.~Krauss-Varban, N.~Omidi, Large-scale hybrid simulations of the magnetotail
  during reconnection, Geophysical Research Letters 22~(23) (1995) 3271--3274.

\bibitem{Wang}
S.~Wang, X.~Xu, Y.-N. Wang, Numerical investigation of ion energy distribution
  and ion angle distribution in a dual-frequency capacitively coupled plasma
  with a hybrid model, Physics of Plasmas 14~(11) (2007) 113501.

\bibitem{chen1984introduction}
F.~F. Chen, Introduction to plasma physics and controlled fusion, Plenum Press,
  New York, 1984.

\bibitem{Mikellides}
I.~G. Mikellides, I.~Katz, R.~R. Hofer, D.~M. Goebel, K.~de~Grys, A.~Mathers,
  Magnetic shielding of the channel walls in a {H}all plasma accelerator,
  Physics of Plasmas 18~(3) (2011) 033501.

\bibitem{Nishikawa2007315}
H.~Nishikawa, A first-order system approach for diffusion equation. {I}:
  Second-order residual-distribution schemes, Journal of Computational Physics
  227~(1) (2007) 315 -- 352.

\bibitem{Nishikawa20103989}
H.~Nishikawa, A first-order system approach for diffusion equation. {II}:
  Unification of advection and diffusion, Journal of Computational Physics
  229~(11) (2010) 3989 -- 4016.

\bibitem{Kajimura:2012aa}
Y.~Kajimura, I.~Funaki, M.~Matsumoto, I.~Shinohara, H.~Usui, H.~Yamakawa,
  Thrust and attitude evaluation of magnetic sail by three-dimensional hybrid
  particle-in-cell code, Journal of Propulsion and Power 28~(3) (2012)
  652--663.

\bibitem{69922}
C.~Gardner, Numerical simulation of a steady-state electron shock wave in a
  submicrometer semiconductor device, Electron Devices, IEEE Transactions on
  38~(2) (1991) 392--398.

\bibitem{Koo}
J.~W. Koo, I.~D. Boyd, Modeling of anomalous electron mobility in {H}all
  thrusters, Physics of Plasmas 13~(3) (2006) 033501.

\bibitem{Komurasaki:1995fk}
K.~Komurasaki, Y.~Arakawa, Two-dimensional numerical model of plasma flow in a
  {H}all thruster, Journal of Propulsion and Power 11~(6) (1995) 1317--1323.

\bibitem{Rognlien1992347}
T.~Rognlien, J.~Milovich, M.~Rensink, G.~Porter, A fully implicit, time
  dependent 2-{D} fluid code for modeling tokamak edge plasmas, Journal of
  Nuclear Materials 196 - 198~(0) (1992) 347 -- 351.

\bibitem{harlow1965numerical}
F.~H. Harlow, J.~E. Welch, et~al., Numerical calculation of time-dependent
  viscous incompressible flow of fluid with free surface, Physics of fluids
  8~(12) (1965) 2182.

\bibitem{Chorin196712}
A.~J. Chorin, A numerical method for solving incompressible viscous flow
  problems, Journal of Computational Physics 2~(1) (1967) 12 -- 26.

\bibitem{Steger1981263}
J.~L. Steger, R.~Warming, Flux vector splitting of the inviscid gasdynamic
  equations with application to finite-difference methods, Journal of
  Computational Physics 40~(2) (1981) 263 -- 293.

\bibitem{Gabutti1983207}
B.~Gabutti, On two upwind finite-difference schemes for hyperbolic equations in
  non-conservative form, Computers \& Fluids 11~(3) (1983) 207 -- 230.

\bibitem{Beam:1978fj}
R.~M. Beam, R.~F. Warming, An implicit factored scheme for the compressible
  {N}avier-{S}tokes equations, AIAA Journal 16~(4) (1978) 393--402.

\bibitem{FUJII:1986xy}
K.~Fujii, S.~Obayashi, Practical applications of new {LU}-{ADI} scheme for the
  three-dimensional {N}avier-{S}tokes computation of transonic viscous flows,
  AIAA Paper 86-0513.

\end{thebibliography}




\end{document}